\documentclass[11pt, a4paper,preprintnumbers,nofootinbib,showpacs]{revtex4}
\usepackage{epsfig}
\usepackage[fleqn,sumlimits,intlimits]{amsmath}
\usepackage{latexsym}
\usepackage[english]{babel}
\usepackage{verbatim}
\usepackage{graphicx}
\usepackage{amsthm}
\usepackage{color}
\usepackage{amssymb}
\usepackage{psfrag}
\usepackage{bm}
\usepackage{bbm}

\def\on#1#2{{\buildrel{\mkern2.5mu#1\mkern-2.5mu}\over{#2}}}

\begin{document}

\title{\Large A Causal Alternative to Feynman's Propagator}

\preprint{ITP-UU-10/43, SPIN-10/36}



\author{Jurjen F. Koksma}
\affiliation{Institute for Theoretical Physics (ITP) \& Spinoza
Institute, Utrecht University, Postbus 80195, 3508 TD Utrecht, The
Netherlands \\ \texttt{\textup{J.F.Koksma@uu.nl}}}

\author{Willem Westra}
\affiliation{Division of Mathematics, The Science Institute, University of Iceland,
Dunhaga 3, 107 Reykjavik, Iceland \\
\texttt{\textup{w.westra@raunvis.hi.is}}}

\begin{abstract}
The Feynman propagator used in the conventional in-out formalism
in quantum field theory is not a causal propagator as wave packets
are propagated virtually instantaneously outside the causal region
of the initial state. We formulate a causal in-out formalism in
quantum field theory by making use of the Wheeler propagator, the
time ordered commutator propagator, which is manifestly causal.
Only free scalar field theories and their first quantization are
considered. We identify the real Klein Gordon field itself as the
wave function of a neutral spinless relativistic particle.
Furthermore, we derive a probability density for our relativistic
wave packet using the inner product between states that live on a
suitably defined Hilbert space of real quantum fields. We show
that the time evolution of our probability density is governed by
the Wheeler propagator, such that it behaves
causally too.
\end{abstract}

\maketitle

\section{Introduction}
\label{Introduction}

Quantum field theory is a very successful theory. The Feynman
rules are very simple to follow but stem from complicated
underlying considerations. For example, in order to compute a
certain transition amplitude, a particular Feynman diagram needs
to be renormalized and computed before it can be related to
observables (a cross section). The standard in-out formalism used
to calculate Feynman diagrams has however two shortcomings:
\begin{itemize}
\item Transition amplitudes in the in-out formalism are not causal;
\item Transition amplitudes in the in-out formalism are not real.
\end{itemize}
The standard in-out formalism is not causal in the following
sense: the Feynman propagator that one uses to calculate diagrams
in perturbation theory is not a causal propagator as it has
non-vanishing support outside the past- and future light cone of a
certain spacetime point $x$. In other words, the Feynman
propagator $\Delta_F(x;x')$ does not vanish when $x$ and $x'$ are
spacelike separated. This
sense of causality is sometimes
referred to as Einstein causality.

The standard in-out formalism is not real in the following sense:
even for real scalar fields the quantum description is complex
since creation operators only create the positive frequency part
of a field and all negative frequency dependence is automatically
projected out. Self-energies are therefore generically complex
valued although we started the computation with real fields.
In other words, the in-out effective action does not generate the
correct quantum corrected equations of motion for the real scalar
field \cite{Park:2010pj, Soussa:2003vn}.

Similar to unitarity \cite{'tHooft:1973pz}, causality is an
ingredient of fundamental importance of relativistic quantum field
theories. Let us be very explicit about the following
point: no one would claim that quantum field theory is not a
causal theory. Expectation values can be expressed in terms of
series of time ordered nested commutators (see e.g.
\cite{Weinberg:2005vy, Koksma:2007uq}), which makes them
manifestly causal. We thus conclude that when calculating
observables using Feynman propagators crucial cancellations must
occur between various diagrams such that the final result is
causal.

The approach which computes true expectation values without
computing scattering amplitudes as an intermediate step is
referred to as the in-in or Schwinger-Keldysh formalism
\cite{Schwinger:1960qe, Keldysh:1964ud, Chou:1984es,
Jordan:1986ug, Calzetta:1986ey, Weinberg:2005vy}, for applications
see \cite{Prokopec:2002uw, Prokopec:2002jn, Prokopec:2003qd,
Koksma:2009wa}. The in-in formalism does not suffer from the
shortcomings of the in-out formalism as mentioned
above\footnote{Although the in-in formalism is perfectly causal,
it is not manifest in all incarnations of the in-in formalism. The
diagrammatic approaches of \cite{Musso:2006pt} and
\cite{Giddings:2010ui} are for instance not based on purely causal
propagators.}. Also, self-mass corrections to real propagators are purely real as
they should. Hence, the interpretation of the self-mass
corrections thus obtained is much more straightforward than in the
in-out formalism.

The main idea that we outline in this paper is how to set up a
real and causal first quantized description in perturbative
quantum field theory in an in-out setting. We only develop the
ingredients necessary to build a perturbative quantum field theory
that meets these requirements, and we postpone the development of
a perturbation theory itself to a future paper. Let us begin by
recalling the first quantized picture in conventional quantum
field theory before we explain where we differ from the standard
approach. Quantum states in the conventional approach to quantum
field theory are complex valued, just as in non-relativistic
quantum mechanics. The reason is very simple to see in second
quantization: a creation operator creates a positive frequency
wave mode out of the vacuum $\hat{a}_{k}^{\dag} |0 \rangle
=|\omega_k,\vec{k} \rangle$, where
$\omega_k=(\vec{k}^2+m^2)^{1/2}$
as usual. Now one can employ the superposition principle such that
the positive frequency contribution $\phi_+(x)$ to the real field
$\phi(x)$ is a quantum mechanical wave function:
\begin{equation}\label{intro1}
\Psi(x) = \phi_+(x) = \int \frac{\mathrm{d}^3 \vec{k}}{(2 \pi)^3 2
\omega_k }\phi(\omega_k,\vec{k}) e^{-i \omega_k t + i \vec{k}
\cdot \vec{x}}\,.
\end{equation}
Clearly, this wave function is complex valued. Furthermore, by
making use of the inner product between fields that live on the
Hilbert space of positive frequency solutions to the Klein Gordon
equation one can straightforwardly find a probability density
that, integrated over its domain of non-zero support, can be
properly normalized to unity. See for example section 3.1 of
\cite{DeWit:1986it} or section 14.2 of \cite{Wald:1984rg} where a
similar interpretation is made. This probability density is the
0th component of the familiar Klein Gordon current given by:
\begin{equation}\label{intro3}
N^{KG}_0(x) = \phi^*_+(x) i \on{\leftrightarrow}{\partial}_t
\phi_+(x)\,,
\end{equation}
The natural question that presents itself when writing down an
equation like (\ref{intro1}) and (\ref{intro3}) is: why is
$\Psi(x)$ not the wave function in quantum field theory? In other
words: why is $\Psi(x)$ not the relativistic generalization of the
wave function of ordinary quantum mechanics? The answer is that
the candidate wave function $\Psi(x)$ defined above does not
behave causally (see e.g.
\cite{Hegerfeldt:1974qu,Hegerfeldt:1998ar}).
Advanced or retarded propagation of the candidate wave function
$\Psi(x)$ is governed by the Feynman propagator, which, as we
emphasized before, is not a causal propagator. It is physically
unacceptable to define a relativistic wave function that would
virtually instantaneously spread outside the past and future light
cone of its region of non-zero support. The Feynman propagator is
however well suited to do perturbation theory as it is time
ordered and hence a Green's function (the solution to the
Schr\"odinger equation is a time ordered evolution operator).
Despite its obvious shortcomings, this is the reason that the
Feynman propagator is so widely used when calculating Feynman
diagrams\footnote{There is an additional problem concerning the
localization of relativistic particles, i.e.: the standard Newton
Wigner position operator suffers from non-locality problems that
would prevent us from localizing a particle on scales smaller than
its Compton wavelength. We will not address this problem here but
present a solution in \cite{Westra1}, which further complements a
consistent first quantized interpretation of quantum field theory.
Additionally, pair creation is often argued to fundamentally limit
the localization of relativistic particles. This argument does not
apply however since we are discussing real fields.}.

We construct \emph{real} quantum states $|\phi\rangle$ that
include both positive and negative frequency contributions $\pm
\omega_k$. Our wave function thus becomes the real field itself:
\begin{equation} \label{intro2}
\Psi(x) = \phi(x) = \phi_+(x) + \phi_-(x) = \int
\frac{\mathrm{d}^3 \vec{k}}{(2 \pi)^3 2 \omega_k }\left[
\phi(\omega_k,\vec{k}) e^{-i \omega_k t + i \vec{k} \cdot \vec{x}}
+ \phi(-\omega_k,\vec{k}) e^{+i \omega_k t + i \vec{k} \cdot
\vec{x}} \right] \,.
\end{equation}
The positive definite inner product between real fields that
includes positive and negative frequency solutions to the Klein
Gordon equation \cite{Henneaux:1982ma, Halliwell:1992nj,
Woodard:1989ac} readily provides us with a positive definite
probability density:
\begin{equation}\label{intro4}
J_0^P(x)=  \frac{1}{2} \left [\phi_+^{\ast}(x) i
\on{\leftrightarrow}{\partial}_t \phi_+(x) - \phi_-^{\ast}(x) i
\on{\leftrightarrow}{\partial}_t \phi_-(x) \right].
\end{equation}
For
real fields, one can easily check that the two currents
(\ref{intro3}) and (\ref{intro4}) are
equal.

What about the propagation of our quantum states, or of the
probability density thus defined? Let us recall again that the
Feynman propagator propagates \emph{complex valued} wave functions
in a manner that is \emph{not causal}, but it is a
\emph{Green's function} such that it can be used in perturbative
quantum field theory. There exists, however, a propagator that
meets all the necessary requirements: it is \emph{real},
\emph{causal} and it is a time ordered \emph{Green's function}
such that we can use it in a perturbative expansion as well. This
propagator is the time ordered commutator propagator, or Wheeler
propagator. The Wheeler propagator is
the real part of the Feynman propagator, or the average of the
advanced and the retarded propagator. Only the Wheeler propagator
propagates the entire real quantum field $\phi(x)$ in a time
ordered manner unlike the Feynman propagator. Ultimately, the idea
is to formally set up a perturbation theory in quantum field
theory using the Wheeler propagator instead of the Feynman
propagator.

The structure of the paper is as follows. In section \ref{First
Quantization including Negative Frequency Contributions}, we
establish our notation and carefully develop a consistent first
quantization in quantum field theory using real states. Also, we
study various propagators to examine how they act on quantum
states. In section \ref{Probability Densities in Quantum Field
Theory}, we consider how probabilities in our new setup of quantum
field theory are defined. Finally, in section \ref{Causal
Propagation of Quantum Fields}, we discuss the role of the Wheeler
propagator and see how it causally propagates probability
densities. Throughout the paper we only consider free, real scalar
field theories to keep the conceptual subtleties clear and
postpone the discussion of interacting field theories to a future
publication.

\section{First Quantization including Negative Frequency Contributions}
\label{First Quantization including Negative Frequency
Contributions}

First quantization attempts to construct a relativistic quantum
field theory for a single particle wave function with a
probabilistic interpretation just as in non-relativistic quantum
mechanics. Here, we do not consider second quantization, where
creation and annihilation operators are introduced to conveniently
describe multi-particle quantum states. The Lagrangian density for
a real scalar field reads:
\begin{equation} \label{Lagrangiandensity1}
{\mathcal L}(\phi,\partial_{\mu}\phi) = -\frac{1}{2} \left
(\eta^{\mu \nu}
\partial_{\mu} \phi \partial_{\nu}\phi + \mu^2 \phi^2
\right) \,,
\end{equation}
where we use the Minkowski metric $\eta^{\mu\nu} =
\mathrm{diag}(-1,1,1,1)$ and where $\mu$ is the inverse of the
reduced Compton wavelength:
\begin{equation} \label{Mass}
\mu = \lambda_{\mathrm{c}}^{-1} = \frac{m c}{\hbar}\,.
\end{equation}
The Klein Gordon equation of motion follows from equation
(\ref{Lagrangiandensity1}) as usual as:
\begin{equation}\label{EoM1}
(\Box - \frac{m^2c^2}{\hbar^2} )\phi(x) =0 \,.
\end{equation}
Here, $\Box$ is the d'Alembertian. Let us Fourier transform our
field as usual as:
\begin{equation}\label{Fouriertransform1}
\phi(x) = \int \frac{\mathrm{d}^4k}{(2 \pi)^4} \tilde{\phi}(k)e^{i
k x} \,,
\end{equation}
where we employed the standard notation $k x =
\eta_{\mu\nu}k^{\mu}x^{\nu}$. Let us
set $\hbar = c = 1$. When
we substitute equation (\ref{Fouriertransform1}) in the Klein
Gordon equation (\ref{EoM1}), we find the usual on shell condition
$(k^0)^2=\vec{k}^2+m^2$,
such that we can write:
\begin{equation} \label{positivenegativefrequency}
\phi(x) = \int \frac{\mathrm{d}^4k}{(2
\pi)^3}\delta(k^2+m^2)\phi(k)e^{i k x}\,.
\end{equation}
We thus see that $\tilde{\phi}(k) = 2\pi\delta(k^2+m^2)\phi(k)$.
We can clearly see that two separate contributions arise if we
were to evaluate the delta function explicitly as we have on shell
solutions with positive and negative frequency:
\begin{equation} \label{positivenegativefrequency2}
\phi_\pm(x) = \int \frac{\mathrm{d}^4k}{(2
\pi)^3}\delta(k^2+m^2)\theta(\pm k^0) \phi(k) e^{i k x} = \int
\frac{\mathrm{d}^3 \vec{k}}{(2 \pi)^3 2 \omega_k }\phi(k_\pm) e^{i
k_\pm x} \,,
\end{equation}
where $\omega_k = (\|\vec{k}\|^2+m^2)^{1/2}$
as usual and we defined the contravariant vector:
\begin{equation} \label{positivenegativefrequency3}
k_\pm  = (\pm \omega_k,\vec{k})\,.
\end{equation}
Throughout the paper we will refer to the $\pm$ indices as
positive and negative frequency indices. If one insists on keeping
a manifestly Lorentz invariant expression, one can leave the delta
function in equation (\ref{positivenegativefrequency2})
unevaluated. We thus have:
\begin{equation} \label{positivenegativefrequency4}
\phi(x)  = \phi_+(x) +\phi_-(x) \,.
\end{equation}
Since we are studying real fields, we have $\phi^{\ast}(-k) =
\phi(k)$ as usual, from which we conclude that $\phi_+^{\ast}(x) =
\phi_-(x)$.

\subsection{Connecting the Schr\"{o}dinger and Klein Gordon equations}

Although the connection between Dirac's equation and
Schr\"odinger's equation is obvious,
one should
appreciate that similar arguments apply to free scalar fields as
well. In non-interacting scalar theories, one can factorize the
Klein Gordon equation (\ref{EoM1}) as follows:
\begin{equation}\label{EoM2}
\left(i \frac{1}{c}\partial_0  +
\sqrt{\mu^2-\vec{\nabla}^2}\right)\left(i \frac{1}{c}\partial_0 -
\sqrt{\mu^2-\vec{\nabla}^2}\right)\phi(x) =0 \,.
\end{equation}
The meaning of the square root of the derivative operator is
defined in momentum space.
In this equation and in the following few we reinstated $\hbar$
and $c$ explicitly again for clarity. Using equation
(\ref{positivenegativefrequency4}) we thus see:
\begin{equation}\label{positivenegativefrequency5}
i \hbar \partial_0 \phi_\pm(x)   =  \pm m
c^2\sqrt{1-\left(\frac{\hbar}{m
c}\right)^2\vec{\nabla}^2}~\phi_\pm(x),
\end{equation}
These equations are sometimes called ``spinless Salpeter''
equations. In the non-relativistic limit where the relative
spatial variations in the field are much smaller than the inverse
Compton wavelength $ \frac{ \vec{\nabla}^2\phi_\pm}{ \phi_\pm} <<
\left(\frac{m c}{\hbar}\right)^2$ one obtains:
\begin{equation}\label{positivenegativefrequency6}
\pm i \hbar \partial_0 \:\phi_\pm(x) =  m c^2 \phi_\pm -
\frac{\hbar^2}{2 m} \vec{\nabla}^2 \phi_\pm - \frac{1}{8}
\frac{\hbar^4}{m^3 c^2} \vec{\nabla}^4\phi_\pm - \ldots \,.
\end{equation}
Clearly, the positive and negative frequency parts of the Klein
Gordon field satisfy a Schr\"{o}dinger equation and its complex
conjugate equation, respectively. We can thus make the following
interpretation: The positive frequency and negative contributions
to a quantum field are wave functions in the non-relativistic
limit. Please note that these considerations only apply to free
theories.

In the literature one can sometimes read confusing statements that
the negative frequency contributions have negative
energy\footnote{These ``negative energy'' contributions were
considered problematic in the early days of quantum field theory
and led Dirac to develop his picture of an infinite sea of
occupied fermionic states which has now become obsolete.}. The
reason for these statements is that the ``Hamiltonian'' appearing
in Schr\"odinger's equation is indeed negative for negative
frequency modes. The appropriate Hamiltonian to consider, however,
is the Klein Gordon Hamiltonian, which is as usual given by:
\begin{flalign}\label{Hamiltonian}
H & = \int \mathrm{d}^4x \partial_0 T^{00}  = \frac{1}{2} \int
\mathrm{d}^3 \vec{x}
 \left((\partial_{0} \phi)^2 + \vec{\nabla}\phi \cdot
\vec{\nabla} \phi  + m^2 \phi^2 \right) \\
& = \frac{1}{2} \int \frac{\mathrm{d}^4k}{(2\pi)^3}
\delta(k^2+m^2)~ |k^0| |\phi(k)|^2 \nonumber\,.
\end{flalign}
We thus conclude that \emph{negative frequency} modes have
\emph{positive energy} as they should\footnote{We would like to
emphasize that we do not introduce fields with negative energy,
contrary to the approaches taken in e.g. \cite{Dirac42, Pauli43,
Linde:1988ws, Kaplan:2005rr, Moffat:2005ip}. We merely include the
Hilbert space of negative frequency solutions to the Klein Gordon
equation into our description of quantum field theory such that
the real field itself can be interpreted as a quantum mechanical
wave function.}.
This point is of course well known in the literature, but we think
it is important to mention it explicitly.

As $\phi_{+}(x)$ satisfies a Schr\"odinger equation in the
non-relativistic regime and because it has positive energies, it
was argued some decades ago that $\phi_{+}(x)$ perhaps was a
promising candidate for the relativistic wave
function\footnote{This is one way to view the
Feynman-St\"{u}ckelberg interpretation of quantum field theory
\cite{Stueckelberg:1941th,Feynman:1948km, Feynman:1949hz}.}. It
was realized however, that these ``wave functions'' would not
spread in a causal manner \cite{Hegerfeldt:1974qu,
Hegerfeldt:1998ar}. We take the following point of view:
\begin{itemize}
\item The entire field $\phi(x)$ and not just its positive frequency part
$\phi_+(x)$ is a wave function in relativistic quantum field
theory. The dynamics of the wave function $\phi(x)$ is governed by the Klein Gordon equation;
\item The probability density follows from the definition of the inner product
on the positive and negative Hilbert spaces.
\end{itemize}
In other words: we include the negative frequency contributions to
a quantum field in what we define to be the wave functions in
quantum field theory. Note that $\hbar$ explicitly appears in the
Klein Gordon equation (\ref{EoM1}),
as is well known, so we can indeed interpret it as a quantum
equation that governs the time evolution of the real wave function
$\phi(x)$. In the next sections, we properly define quantum
states, study how they are propagated by various propagators and
finally, we associate probabilities to the inner product of a
quantum state with itself and study how this probability density
is propagated in time.

\subsection{Quantum States in Quantum Field Theory}

Let us firstly recall how wave packets are constructed in the
conventional approach to quantum field theory. Here, the crucial
point is that a quantum state is derived from the positive
frequency contribution to a quantum field only. Those states are
therefore complex valued. For example,
equation (4.65) of Peskin and Schroeder \cite{Peskin:1995ev}
reads:
\begin{equation}\label{ordinaryquantumstate}
|\phi_{\mathrm{PS}} \rangle = \int \frac{\mathrm{d}^3 \vec{k}}{(2
\pi)^3 \sqrt{2\omega_k}}\phi(\vec{k})| \vec{k} \rangle\,,
\end{equation}
where the subscript ``PS'' refers to Peskin and Schroeder. In
order to better illustrate the point we would like to make, we
modify this equation slightly to:
\begin{equation}\label{ordinaryquantumstate2}
|\phi_{\mathrm{PS}} \rangle \rightarrow \int \frac{\mathrm{d}^3
\vec{k}}{(2 \pi)^3 2\omega_k }\phi(\omega_k,\vec{k})|\omega_k,
\vec{k} \rangle\,.
\end{equation}
Equation (\ref{ordinaryquantumstate}) and
(\ref{ordinaryquantumstate2}) differ in two respects:
normalization\footnote{The Lorentz invariant normalization of
quantum states appears for example in \cite{Brown:1992db} formula
3.4.2.} and a notational difference as we explicitly include the
$k^0$ dependence. Unlike Peskin and Schroeder, we insist on a
Lorentz invariant definition for the states. One can express any
state in a manifestly Lorentz invariant way as:
\begin{equation}\label{ordinaryquantumstate3}
|\tilde{\phi} \rangle = \int  \mathrm{d} (LIPS)\
\tilde{\phi}(k^0,\vec{k})|k^0,\vec{k} \rangle\,,
\end{equation}
where the Lorentz invariant phase space (LIPS) element for one
particle is in the standard approach to quantum field theory given
by\footnote{See for instance Ryder's formula (4.4)
\cite{Ryder:1985wq}.}:
\begin{equation}\label{ordinaryquantumstate4}
\mathrm{d}(LIPS) = \frac{\mathrm{d}^4
k}{(2\pi)^3}\delta(k^2+m^2)\theta(k^0) =
\mathrm{d}k^0\frac{\mathrm{d}^3 \vec{k}}{(2 \pi)^3 2\omega_k
}\delta(k^0-\omega_k)\,.
\end{equation}
It is clear from equations (\ref{ordinaryquantumstate2}) and
(\ref{ordinaryquantumstate4}) above that in the Peskin and
Schroeder state (\ref{ordinaryquantumstate}) only positive
frequencies $+\omega_k$ contribute\footnote{The exclusion of
negative frequency states is essentially due to Feynman and
St\"{u}ckelberg \cite{Stueckelberg:1941th,Feynman:1948km,
Feynman:1949hz} who understood negative frequency states as
Hermitian conjugate positive frequency states in the context of
charged fields.}:
\begin{equation}\label{ordinaryquantumstate5}
|\phi_{\mathrm{PS}} \rangle  = |\phi_+ \rangle \,,
\end{equation}
where we added a + sign to the right hand side of this equation to
make this explicit.

Now let us discuss how we define quantum states in our new setup
of quantum field theory. Given the fact that we can interpret a
real scalar field as a wave function in relativistic quantum
mechanics, we can define a Lorentz invariant state for our field
as follows:
\begin{subequations}
\label{fieldstates1}
\begin{flalign}
& |\phi \rangle = \int \frac{\mathrm{d}^4k}{(2
\pi)^3}\delta(k^2+m^2)\phi(k)| k \rangle \label{fieldstates1a}\\
& \langle \phi | = \int \frac{\mathrm{d}^4k}{(2
\pi)^3}\delta(k^2+m^2)\phi^*(k)\langle k| \label{fieldstates1b}\,.
\end{flalign}
\end{subequations}
Clearly, a quantum state is a superposition of various wave modes
$| k \rangle$ with a certain amplitude $\phi(k)$ as usual. If one
evaluates the delta function the positive and negative frequency
modes appear again such that the state of a field is a
superposition of positive and negative frequency states:
\begin{equation} \label{fieldstates2}
|\phi \rangle  = |\phi_+\rangle  +|\phi_-\rangle  \,,
\end{equation}
where:
\begin{equation}\label{fieldstates3}
|\phi_\pm\rangle =
\int \frac{\mathrm{d}^4k}{(2\pi)^3}\delta(k^2+m^2)\theta(\pm k^0) \phi(k)| k \rangle
= \int \frac{\mathrm{d}^3 \vec{k}}{(2 \pi)^3
2\omega_k }\phi(k_\pm)| k_\pm \rangle \,,
\end{equation}
and where:
\begin{equation}\label{fieldstates4}
|k_\pm\rangle= | \pm \omega_k,\vec{k}\rangle\,.
\end{equation}
It is of fundamental importance to realize that, unlike the
conventional approach taken in the quantum field theoretical
literature as in equation (\ref{ordinaryquantumstate}), our
quantum state (\ref{fieldstates1}) includes the negative frequency
contributions. This makes our quantum state real.

The inner product between momentum basis bras and state kets is
naturally obtained by demanding that:
\begin{equation}\label{fieldstates5}
\langle k_\pm | \phi \rangle = \phi_ \pm(k)\,,
\end{equation}
such that:
\begin{equation}\label{momentuminnerproduct}
 \langle k_a | k_b' \rangle =  (2\pi)^3 2 \omega_k \delta^{(3)}(\vec{k}-\vec{k}') \delta_{ab} \,,
\end{equation}
where $a$ and $b$ are the frequency indices $\pm$. This relation
was first introduced by Hennaux and Teitelboim, who presented a
worldline discussion of the supersymmetric particle
\cite{Henneaux:1982ma}. We can now introduce spacetime kets as the
covariant on shell Fourier transform of the momentum kets:
\begin{equation}\label{spacetimebraket1}
|x \rangle =\int \frac{\mathrm{d}^4k}{(2
\pi)^3}\delta(k^2+m^2)e^{- ik x}| k\rangle = |x_+ \rangle + |x_-
\rangle\,,
\end{equation}
where:
\begin{subequations}
\label{spacetimebraket2}
\begin{flalign}
& |x_{\pm} \rangle = \int \frac{\mathrm{d}^4k}{(2
\pi)^3}\delta(k^2+m^2)\theta(\pm k^0) e^{- ik x}| k\rangle = \int
\frac{\mathrm{d}^3 \vec{k}}{(2 \pi)^3 2\omega_k
}e^{-ik_\pm x}| k_\pm \rangle \label{spacetimebraket2a}\\
& \langle x_\pm | = \int \frac{\mathrm{d}^3 \vec{k}}{(2 \pi)^3
2\omega_k }e^{ik_\pm x}\langle k_\pm|\label{spacetimebraket2b}\,.
\end{flalign}
\end{subequations}
The sign in both exponents is chosen to give the standard sign for
the Fourier transform in equation (\ref{Fouriertransform1}) as we
show below in equation (\ref{wavemodes1}). Equation
(\ref{spacetimebraket1}) merits another remark. If we compare this
equation for $|x\rangle$ with equation (\ref{fieldstates1a}) for
$|\phi\rangle$, we can identify $\phi(k) \Leftrightarrow e^{- ik
x}$. In other words: the state $|x\rangle$ corresponds to a
superposition of plane wave modes, i.e.: a non-normalizable
``wavepacket'' of plane waves. The inverse Fourier relations are given by:
\begin{subequations}
\label{spacetimebraket3}
\begin{flalign}
& |k_\pm \rangle = \mp \int \mathrm{d}^3 \vec{x} e^{ik_\pm x} i
\on{\leftrightarrow}{\partial}_0 | x_\pm \rangle \label{spacetimebraket3a}\\
& \langle k_\pm | = \pm \int \mathrm{d}^3 \vec{x} e^{-ik_\pm x} i
\on{\leftrightarrow}{\partial}_0 \langle x_\pm |
\label{spacetimebraket3b}\,.
\end{flalign}
\end{subequations}
Here, we defined a time derivative that acts in the following way
on two test functions $f(x)$ and $g(x)$:
\begin{equation}\label{timederivative}
f(x) \on{\leftrightarrow}{\partial}_0  g(x) = f(x)
\partial_0  g(x) - g(x)
\partial_0  f(x)\,.
\end{equation}
Having carefully defined the spacetime states, we find:
\begin{equation}\label{FieldWavefuncproject}
\phi(x) =\phi_+(x) + \phi_-(x)= \langle x_+ | \phi\rangle +
\langle x_- | \phi\rangle = \langle x | \phi \rangle \,.
\end{equation}
We can furthermore derive:
\begin{equation}\label{FieldWavefuncproject2}
|\phi_\pm \rangle = \mp \int \mathrm{d}^3 \vec{x} \phi_\pm (x) i
\on{\leftrightarrow}{\partial}_0 | x_\pm \rangle \,.
\end{equation}
Using (\ref{spacetimebraket2}) and the positive definite momentum
space inner product \eqref{momentuminnerproduct} one easily
derives:
\begin{subequations}
\label{wavemodes1}
\begin{flalign}
& \langle x_\pm | k_\pm \rangle = e^{ik_\pm x}\label{wavemodes1a}\\
& \langle k _\pm | x_ \pm \rangle = e^{-ik_\pm x}
\label{wavemodes1b} \\
& \langle x_\pm | k_\mp \rangle = 0 \label{wavemodes1c} \\
& \langle k_\pm | x_\mp \rangle = 0 \label{wavemodes1d} \,.
\end{flalign}
\end{subequations}
%

\subsection{Inner Products and the Identity Operator}

We can also compute various position space inner products. The
inner product between the positive and negative frequency modes
separately yields the two Wightman functions:
\begin{equation}\label{Wightmanfunctions}
\Delta^\pm(x-x') = \pm i \langle x_\pm| x_\pm' \rangle   = \pm i
\int \frac{\mathrm{d}^4k}{(2 \pi)^3} \delta(k^2+m^2) \theta(\pm
k^0)e^{ik(x-x')} = \pm i \int \frac{\mathrm{d}^3\vec{k}}{(2 \pi)^3
2\omega_k} e^{ik_\pm(x-x')} \,.
\end{equation}
The $\pm i $ in the definition of the Wightman functions becomes
apparent later when we discuss propagators. Moreover, it is
important to stress that our $\pm$ frequency indices are very
different from the Schwinger-Keldysh indices often used in the
in-in formalism in out-of-equilibrium quantum field theory. We
also find:
\begin{equation} \label{spacetimeinnerproduct}
\langle x_\pm| x_\mp' \rangle =0\,.
\end{equation}
The Wightman functions furthermore obey:
\begin{subequations}
\label{Wightmanfunctionsproperties1}
\begin{flalign}
\Delta^+(x-x')  &= - \Delta^-(x'-x)
\label{Wightmanfunctionsproperties1a} \\
\left[ \Delta^\pm(x-x') \right]^{\ast}  & = \Delta^\mp(x-x')
\label{Wightmanfunctionsproperties1b} \,.
\end{flalign}
\end{subequations}
The positive definite inner product between two arbitrary states
follows from equation (\ref{fieldstates1}) as:
\begin{flalign}\label{inner productfields}
\langle \chi | \phi \rangle =& \int \frac{\mathrm{d}^4k}{(2
\pi)^3}\delta(k^2+m^2)\chi^*(k)\phi(k) = \int
\frac{\mathrm{d}^3\vec{k}}{(2 \pi)^3 2\omega_k} \left[\langle \chi
| k_+ \rangle \langle k_+ |\phi \rangle +\langle \chi| k_- \rangle
\langle k_- |\phi\rangle \right] \\
=& \int \frac{\mathrm{d}^4k}{(2 \pi)^3}\delta(k^2+m^2)\langle
\chi| k \rangle \langle k | \phi\rangle \nonumber \,.
\end{flalign}
From the equation above we read off a Lorentz covariant identity
operator:
\begin{equation}\label{identitymomentum1}
\hat{I} = \int \frac{\mathrm{d}^4k}{(2 \pi)^3} \delta(k^2+m^2) |
k\rangle \langle k| = \hat{I}_+ + \hat{I}_- \,,
\end{equation}
where:
\begin{equation}\label{identitymomentum2}
\hat{I}_\pm = \int \frac{\mathrm{d}^3\vec{k}}{(2 \pi)^3 2\omega_k}
| k_\pm\rangle \langle k_\pm| \,,
\end{equation}
which we can easily verify by for example applying it to $|
k_\pm\rangle $. To derive the position space identity operator let
us consider objects like $\langle x_\pm | x_\pm' \rangle$ derived
in equation (\ref{Wightmanfunctions}). Inserting the following
operator leaves these functions unchanged:
\begin{equation}\label{identityposition}
\hat{I}_\pm = \pm \int \mathrm{d}^3 \vec{x} | x_\pm\rangle i
\on{\leftrightarrow}{\partial}_0 \langle x_\pm| \,,
\end{equation}
such that the position space identity operator on the total space
of solutions is given by:
\begin{equation}\label{identityposition2}
\hat{I} = \int \mathrm{d}^3 \vec{x} | x_+\rangle i
\on{\leftrightarrow}{\partial}_0 \langle x_+|- | x_-\rangle i
\on{\leftrightarrow}{\partial}_0 \langle x_-|\,.
\end{equation}
%

\subsection{Propagators, Propagation Laws and Composition Laws}

In this subsection, we derive various propagators for our quantum
field. Intuitively, it is important to realize that propagators
only propagate a state but do not contain information about the
state itself. In other words: the wave mode $|k\rangle$ of a
certain state has an amplitude $\phi(k)$ as in equation
(\ref{fieldstates1a}). Wave packets can thus be constructed by the
principle of superposition. Propagators do not contain any of this
information. The amplitude of a certain wave mode $|k\rangle$ for
a propagator is just the plane wave $\phi(k) = e^{-ikx}$. A
propagation law captures how a propagator acts on a certain wave
function, whereas a composition law captures how a propagator acts
on another propagator.

\subsubsection{Wightman propagators}

We have already encountered the two Wightman functions in equation
(\ref{Wightmanfunctions}), given by: \mbox{$ \Delta^\pm(x-x') =
\pm i \langle x_\pm| x_\pm' \rangle $}. What we discuss here is
how they act on the wave functions. By making use of our identity
operator (\ref{identityposition}) one can straightforwardly show
that:
\begin{equation}\label{propagationlaw1}
\phi_{\pm}(x) = \pm i \int \mathrm{d}^3 \vec{x}'  \langle x_{\pm}
| x'_{\pm} \rangle  \on{\leftrightarrow}{\partial}_{t'}
\phi_{\pm}(x') = \int \mathrm{d}^3 \vec{x}' \Delta^{\pm}(x-x')
\on{\leftrightarrow}{\partial}_{t'} \phi_{\pm}(x')\,.
\end{equation}
The propagators for positive and negative frequency wave functions
are the positive and negative frequency Wightman propagators,
respectively. Clearly, a wave function at a time $t'$ is
transformed to a wave function at another time $t$, where at the
moment $t$ can be either earlier or later than $t'$. It is
important to realize that information about the state, i.e.: the
specific superposition of wave modes in (\ref{fieldstates1}), is
always contained in $\phi_{\pm}$, whereas the propagation of a
state is governed by the Wightman functions. Wightman propagators
can also act on the complex conjugate wave functions:
\begin{equation}\label{propagationlaw2}
\phi_{\pm}^*(x) =\pm i \int \mathrm{d}^3 \vec{x}' \phi_{\pm}^*(x')
\on{\leftrightarrow}{\partial}_{t'}\langle x'_{\pm} | x_{\pm}
\rangle =  \int \mathrm{d}^3 \vec{x}' \phi_{\pm}^*(x')
\on{\leftrightarrow}{\partial}_{t'} \Delta^{\pm}(x'-x) \,.
\end{equation}
Here, we have made use of equation (\ref{Wightmanfunctions}).
Moreover, we can easily see that:
\begin{equation}
\int \mathrm{d}^3 \vec{x}' \Delta^{\pm}(x-x')
\on{\leftrightarrow}{\partial}_{t'} \phi_{\mp}(x') = 0
\label{propagationlaw3}
\end{equation}
The Wightman propagators satisfy the following composition laws:
\begin{equation}\label{compositionlaw1}
\int \mathrm{d}^3 \vec{x}'' \Delta^\pm(x-x'')
\on{\leftrightarrow}{\partial}_{\!t''}  \Delta^\pm(x''-x') =
\Delta^\pm(x-x')\,,
\end{equation}
which we prove straightforwardly by making use of the integral
representation (\ref{Wightmanfunctions}). It will come as no
surprise that the mixed composition laws for the Wightman
propagators vanish:
\begin{equation}\label{compositionlaw2}
\int \mathrm{d}^3\vec{x}''  \Delta^\pm(x-x'')
\on{\leftrightarrow}{\partial}_{\!t''} \Delta^\mp(x''-x')=0 \,.
\end{equation}
%

\subsubsection{Feyman propagator}

The Feynman propagator ensures that the ``right to left''
direction of propagation in equation (\ref{propagationlaw1})
coincides with the direction of increasing time:
\begin{subequations}
\label{Feynmanpropagator}
\begin{flalign}
& \theta(t-t')\phi_+(x) = \int \mathrm{d}^3 \vec{x}'  \Delta_F(x-x')  \on{\leftrightarrow}{\partial}_{t'} \phi_+(x') \label{Feynmanpropagatora}\\
& \theta(t'-t)\phi_+^*(x) = \int \mathrm{d}^3 \vec{x}'
\phi_+^*(x') \on{\leftrightarrow}{\partial}_{t'} \Delta_F(x-x')\,,
\label{Feynmanpropagatorb}
\end{flalign}
\end{subequations}
where the Feynman propagator is the time ordered positive
frequency Wightman propagator:
\begin{flalign}\label{Feynmanpropagator2}
\Delta_F(x-x') =& \, i T[\langle x_+ | x_+' \rangle] =  i
\theta(t-t') \langle x_+ | x_+' \rangle + i \theta(t'-t) \langle
x'_+ | x_+ \rangle \\
=& \, i \! \int \! \frac{\mathrm{d}^4k}{(2 \pi)^3} \delta(k^2+m^2)
\theta(k^0)T\left[ e^{ik(x-x')}\right] \nonumber \,.
\end{flalign}
Here, $T$ denotes the time ordering symbol. This form shows
explicitly that the Feynman propagator describes the on shell time
ordered propagation of the positive frequency modes of a field. In
other words, the Feynman propagator is a retarded propagator for
positive frequency modes. Also, we see that the Feynman propagator
is an advanced propagator for the complex conjugates of positive
frequency modes. Since we know that \mbox{$\phi^{\ast}_+(x) =
\phi_-(x)$,} we conclude from equation (\ref{Feynmanpropagatorb})
that the Feynman propagator is an advanced propagator of the
negative frequency contributions too. We can bring equation
(\ref{Feynmanpropagator2}) in the conventional form by also
Fourier transforming the Heaviside step functions:
\begin{flalign}\label{Feynmanpropagator3}
\Delta_F(x-x')  =& \,\, i \! \int \frac{\mathrm{d}E}{2\pi} \int
\frac{\mathrm{d}^3\vec{k}}{(2 \pi)^32\omega_k }\left[
\frac{i}{E+i\epsilon}e^{-iE(t-t')+ik_+ (x-x')} -
\frac{i}{E-i\epsilon}e^{-iE(t-t') -ik_+( x-x')}\right]\nonumber \\
=& \int \frac{\mathrm{d}^4k}{(2 \pi)^4} \frac{1}{k^2 + m^2 -
i\epsilon}e^{ik(x-x')}\,,
\end{flalign}
where we shifted the $E$ integration appropriately in the second
line. Alternatively, we can rewrite the equation above as:
\begin{equation}\label{Feynmanpropagator4}
\Delta_F(x-x') = P \left[\int \frac{\mathrm{d}^4k}{(2 \pi)^4}
\frac{1}{k^2 + m^2}e^{ik(x-x')} \right] + i \pi \int
\frac{\mathrm{d}^4k}{(2 \pi)^4}\delta(k^2+m^2)e^{ik(x-x')}\,,
\end{equation}
where we have made use of the Sokhotski-Plemelj or
Sokhotsky-Weierstrass formula, also known as Dirac's rule:
\begin{equation}\label{Diracsrule}
\frac{1}{\mu \pm i\varepsilon} =P \frac{1}{\mu} \mp i \pi
\delta(\mu)\,,
\end{equation}
where $P$ denotes the Cauchy principal value. The anti-Feynman or
anti-time ordered propagator ($\overline{T}$) is now given by:
\begin{flalign}\label{antiFeynmanpropagator1}
\Delta_{AF}(x-x') & = i \overline{T} [\langle x_+ | x_+' \rangle] =
i \theta(t'-t) \langle x_+ | x_+' \rangle + i \theta(t-t') \langle
x'_+ | x_+ \rangle \\
& = \int \frac{\mathrm{d}^4k}{(2 \pi)^4}
\frac{1}{k^2 + m^2 + i\epsilon}e^{ik(x-x')} \nonumber\,.
\end{flalign}

Let us now discuss the various composition laws for the Feynman
propagator. Firstly, a Feynman propagator can act on another
Feynman propagator:
\begin{equation}\label{compositionlaw3}
\int \mathrm{d}^3\vec{x}'' \Delta_F(x-x'')
\on{\leftrightarrow}{\partial}_{\!t''} \Delta_F(x''-x') =
\theta(t-t'') \theta (t''-t') \Delta^+(x-x')+\theta(t''-t)
\theta(t'-t'')\Delta^-(x-x') \,,
\end{equation}
where it is crucially important to make use of the integral
representation of the Feynman propagator in equation
(\ref{Feynmanpropagator3}) in order to take the distributional
character of the propagator properly into account. If we time
order the expression above, i.e.: we require the latest to the
left convention, we find the composition law for the Feynman
propagator:
\begin{flalign}\label{compositionlaw4}
& T \left[ \int \mathrm{d}^3\vec{x}'' \Delta_F(x-x'')
\on{\leftrightarrow}{\partial}_{\!t''} \Delta_F(x''-x')\right]
\\
& =  \theta(t-t')\int \mathrm{d}^3\vec{x}'' \Delta_F(x-x'')
\on{\leftrightarrow}{\partial}_{\!t''} \Delta_F(x''-x') +
\theta(t'-t) \int \mathrm{d}^3\vec{x}'' \Delta_F(x'-x'')
\on{\leftrightarrow}{\partial}_{\!t''} \Delta_F(x''-x) \nonumber \\
& = \Delta_F(x-x') \Big|_{
\begin{subarray}{l}
t'' \neq t \\
t'' \neq t'
\end{subarray}}
+\frac{1}{2} \Delta_F(x-x') \Big|_{
\begin{subarray}{l}
t'' \neq t \\
t'' = t'
\end{subarray}}
+\frac{1}{2} \Delta_F(x-x') \Big|_{
\begin{subarray}{l}
t'' = t \\
t'' \neq t'
\end{subarray}} \nonumber\,.
\end{flalign}
Also, we can study how a Feynman propagator or anti-Feynman
propagator acts on our Wightman functions. By making use of
equation (\ref{Feynmanpropagator3}) and of several complex contour
integrations one can show:
\begin{subequations}
\label{compositionlaw5}
\begin{flalign}
& \int \mathrm{d}^3\vec{x}'' \Delta_F(x-x'')
\on{\leftrightarrow}{\partial}_{\!t''} \Delta^{\pm}(x''-x') = \pm
\theta(\pm t \mp t'') \Delta^{\pm}(x-x')
\label{compositionlaw5a}\\
& \int \mathrm{d}^3\vec{x}'' \Delta_{AF}(x-x'')
\on{\leftrightarrow}{\partial}_{\!t''} \Delta^{\pm}(x''-x') = \mp
\theta(\mp t \pm t'') \Delta^{\pm}(x-x') \label{compositionlaw5b}
\,.
\end{flalign}
\end{subequations}
Apart from being a propagator, the Feynman propagator is also a
Green's function for the Klein Gordon equation. By making use of
equation (\ref{Feynmanpropagator3}) one can verify that:
\begin{equation}\label{GreensfunctionFeynman}
(-\partial^2 + m^2)\Delta_F(x-x') = \delta^{(4)}(x-x')\,,
\end{equation}
such that the solution of $(-\partial^2 + m^2)\phi_+(x) =
\tilde{J}(x)$ is:
\begin{equation}\label{GreensfunctionFeynman2}
\phi_+(x)= \tilde{\phi}^0(x)+\int \mathrm{d}^4x' \Delta_F(x-x')
\tilde{J}(x') \,.
\end{equation}
Here, $\tilde{\phi}^0(x)$ is a complex valued solution to the
homogeneous Klein Gordon equation, and $\tilde{J}$ is a complex
valued source. Note that the Feyman propagator is complex valued
and is therefore certainly not the Green's function of the real
scalar field $\phi$. It can however be the Green's function of
e.g. the positive frequency contribution to our real field.

\subsubsection{Commutator Propagator}

So far, we have introduced the two Wightman and Feynman
propagators. The former propagate the positive and negative
frequency contributions to a real field, whereas the latter act as
retarded and advanced propagators for these contributions,
respectively. There is, however, another propagator worth
discussing. It is the commutator propagator and it propagates the
entire field, i.e.: both its positive and negative frequency
contributions:
\begin{equation}\label{commutatorpropagator1}
\phi(x) = \int \mathrm{d}^3 \vec{x}' \Delta_C(x-x')
\on{\leftrightarrow}{\partial}_{t'} \phi(x')
\end{equation}
where the commutator propagator $\Delta_C(x-x')$ is given by:
\begin{equation}\label{commutatorpropagator2}
\Delta_C(x-x') = i \eta^{ab}\langle x_a| x_b' \rangle= i \langle
x_+| x_+' \rangle - i \langle x_-| x_-' \rangle=\Delta^+(x-x')+
\Delta^-(x-x')\,.
\end{equation}
Its integral representation reads:
\begin{equation}\label{commutatorpropagator3}
\Delta_C(x-x') =  i \int \frac{\mathrm{d}^4k}{(2 \pi)^3}
\delta(k^2+m^2)\epsilon(k^0) e^{ik(x-x')}\,,
\end{equation}
where $\epsilon(k^0)$ is sign function as before. It is important
to realize that the commutator propagator satisfies the following
relations:
\begin{subequations}
\label{commutatorpropagator4}
\begin{flalign}
\left. \Delta_C(x-x') \right|_{t=t'} & = 0
\label{commutatorpropagator4a}\\ \left. \partial_t
\Delta_C(x-x')\right|_{t=t'} & = \delta^3(\vec{x}-\vec{x}')
\label{commutatorpropagator4b}\,.
\end{flalign}
\end{subequations}
These identities are important for the following reason:
\begin{flalign} \label{commutatorpropagator5}
\lim_{t \rightarrow t'} \phi(x) &= \int \mathrm{d}^3 \vec{x}'
\Delta_C(x-x') \partial_{t'} \phi(x') \Big|_{t=t'} - \int
\mathrm{d}^3 \vec{x}' \phi(x')
\partial_{t'} \Delta_C(x-x') \Big|_{t = t'} \nonumber \\
&= \phi(\vec{x},t') \,.
\end{flalign}
Thus at time coincidence no propagation of the field has occurred.
The commutator propagator satisfies the following composition law:
\begin{equation}\label{compositionlaw6}
\Delta_C(x-x') = \int \mathrm{d}^3\vec{x}' \Delta_C(x-x'')
\on{\leftrightarrow}{\partial}_{\!t''}  \Delta_C(x''-x')\,.
\end{equation}
%

\subsubsection{Wheeler Propagator}

Clearly, the commutator propagator is not a Green's function as it
does not satisfy the inhomogeneous Klein Gordon equation. Let us
therefore define the Wheeler propagator as the time ordered
commutator propagator:
\begin{equation}\label{Wheelerpropagator}
\Delta_W(x-x') = \frac{1}{2} T [\Delta_C(x-x')] =
\frac{\epsilon(t-t')}{2} \Delta_C(x-x') \,.
\end{equation}
The last identity follows from the anti-symmetry of the
commutator. The Wheeler propagator is thus symmetric in its
arguments\footnote{This propagator appears in a classical context
in \cite{Damour:2001bu,Damour:2008ji}.}. We follow
\cite{Bollini:1998hj} in calling the time ordered commutator the
Wheeler propagator\footnote{We do not agree however with the
statement made in \cite{Bollini:1998hj} that the Wheeler
propagator lacks on shell propagation, as is clear from equation
\eqref{Wheelerpropagator3}.} as it appears implicitly in Wheeler
and Feynman's absorber theory \cite{Wheeler:1945ps}. Wheeler's two
point function is also the average of the advanced and retarded
propagator \cite{Birrell:1982ix, Itzykson:1980rh}:
\begin{equation}
\Delta_W(x-x') = \frac{1}{2}\left(\Delta_A(x-x') +
\Delta_R(x-x')\right)\,,
\end{equation}
where:
\begin{subequations}
\label{AdvancedRetarded1}
\begin{flalign}
& \Delta_R(x-x') = \theta(t-t')\Delta_C(x-x') \label{AdvancedRetarded1a}\\
& \Delta_A(x-x') = - \theta(t'-t)\Delta_C(x-x')
\label{AdvancedRetarded1b}\,.
\end{flalign}
\end{subequations}
Moreover, the Wheeler propagator satisfies the following
propagation law:
\begin{equation} \label{Wheelerpropagationlaw}
\frac{\epsilon(t-t')}{2} \phi(x) =  \int \mathrm{d}^3 \vec{x}'
\Delta_W(x-x') \on{\leftrightarrow}{\partial}_{t'} \phi(x') \,.
\end{equation}
Hence, like the commutator propagator, the Wheeler propagator
propagates the entire real quantum field. Note that this is
natural since Wheeler's two point function is real unlike the
Feynman propagator. The Wheeler propagator has the following
integral representations:
\begin{flalign}\label{Wheelerpropagator3}
\Delta_W(x-x')  &=  \frac{i}{2}  \int \frac{\mathrm{d}^4k}{(2
\pi)^3} \delta(k^2+m^2)\epsilon(k^0)
T\left[e^{ik(x-x')}\right] \nonumber \\
& = \frac{i}{2}  \int \frac{\mathrm{d}^4k}{(2 \pi)^3}
\delta(k^2+m^2)\epsilon(k^0)\epsilon(t - t') e^{ik(x-x')}\nonumber
 \\ &= \frac{1}{2} \int \frac{\mathrm{d}^4k}{(2
\pi)^4}\left[ \frac{1}{k^2+m^2+i \epsilon} +  \frac{1}{k^2+m^2- i
\epsilon}\right]e^{ik(x-x')} \,.
\end{flalign}
where on the second line we used the Fourier representation of the
Heaviside step function. The Wheeler propagator is thus given by
the real part of the Feynman propagator:
\begin{flalign}\label{Wheelerpropagator2}
\Delta_W(x-x') & = \frac{1}{2}\left(\Delta_F(x-x') +
\Delta_{AF}(x-x') \right) =  P \left[ \int \frac{\mathrm{d}^4k}{(2
\pi)^4}\frac{1}{k^2+m^2} e^{ik(x-x')} \right] \nonumber \\
& = \Re\Delta_F(x-x') \,.
\end{flalign}
Here, we used Dirac's rule in equation (\ref{Diracsrule}) again.
Clearly, this would equally well allow us to refer to the Wheeler
propagator as the ``principal part propagator'', see for instance
\cite{Greiner:1996zu}. Let us now discuss the composition law for
the Wheeler propagator. A Wheeler propagator acts on another
Wheeler propagator as follows:
\begin{equation}\label{compositionlaw10}
\int \mathrm{d}^3\vec{x}'' \Delta_W(x-x'')
\on{\leftrightarrow}{\partial}_{\!t''} \Delta_W(x''-x') =
\frac{1}{4} \Delta_C(x-x') \epsilon(t-t'') \epsilon (t''-t')\,.
\end{equation}
Here, we made frequent use of equation (\ref{compositionlaw5}). Like
the composition law of the Feynman propagator, we should time
order the expression above. We thus derive:
\begin{equation}\label{compositionlaw11}
T \left[ \int \mathrm{d}^3\vec{x}'' \Delta_W(x-x'')
\on{\leftrightarrow}{\partial}_{\!t''} \Delta_W(x''-x') \right] =
\frac{1}{2} \Delta_W(x-x') \epsilon(t-t'') \epsilon (t''-t')\,.
\end{equation}
We can recognize this equation as a composition law if we require
$t <t''<t' ~\cup~ t'<t''<t$:
\begin{equation}\label{compositionlaw11b}
 \Delta_W(x-x') = 2T \left[ \int \mathrm{d}^3\vec{x}'' \Delta_W(x-x'')
\on{\leftrightarrow}{\partial}_{\!t''} \Delta_W(x''-x') \right]_{t <t''<t' ~\cup~ t'<t''<t}.
\end{equation}
The Wheeler propagator is not only a propagator, it is also a
Green's function of the Klein Gordon equation because of the time
ordering. One can again easily see that:
\begin{equation}\label{GreensfunctionWheeler}
(-\partial^2 + m^2)\Delta_W(x-x') = \delta^{(4)}(x-x')\,,
\end{equation}
where we have made use of equation (\ref{Wheelerpropagator3}). The
solution of $(-\partial^2 + m^2)\phi(x) = J(x)$ is thus:
\begin{equation}\label{GreensfunctionWheeler2}
\phi(x)= \phi^0(x)+\int \mathrm{d}^4x' \Delta_W(x-x') J(x') \,.
\end{equation}
Here, $\phi^0(x)$ is a real solution to the homogeneous Klein
Gordon equation. Finally, note that the Wheeler propagator is the
Green's function of the real scalar field $\phi$, and not just of
the complex valued positive frequency contribution $\phi_+$ as the
Feyman propagator is.

\section{Probability Densities in Quantum Field Theory}
\label{Probability Densities in Quantum Field Theory}

Given our wave function interpretation of the real quantum field
$\phi$ as presented above, it is natural to define probability
densities analogous to quantum mechanics. In quantum mechanics,
the absolute value squared of a quantum mechanical wave function
represents a probability distribution function of a single quantum
mechanical particle, whose dynamics is governed by Schr\"odinger's
equation. The absolute value squared of a quantum mechanical wave
function can be used as a probability measure because firstly it
is positive and, secondly, it can be normalized to unity when
integrated over its domain. The natural candidate for a
probability density in relativistic quantum mechanics stems from
the inner product between a quantum state with itself as we show
shortly. The probability densities thus obtained are positive and
can be normalized to unity too.

In the conventional approach to quantum field theory this natural
generalization of quantum mechanics is problematic as only the
positive frequency modes are included in wave packets and the
negative frequency modes do not contribute. The propagation of
these wave packets is, as we have seen above, governed by the
Feynman propagator. The wave packets therefore spread
non-causally. We solve this problem by properly including the
negative frequency contributions in the wave packets such that
their propagation is mediated by Wheeler's propagator. We
therefore have a normalized probability density at our disposal
that spreads in a causal manner. This allows us to build a
consistent first quantized picture in quantum field theory that
naturally generalizes quantum mechanics to a relativistic setting.

\subsection{A Normalized Probability Density and the Hadamard Propagator}

In our setup of quantum field theory we require:
\begin{equation}\label{NormalizationStates}
\frac{1}{2} \langle \phi | \phi \rangle \equiv 1 \,.
\end{equation}
Inserting the identity operator from equation
(\ref{identityposition2}) yields:
\begin{flalign}\label{NormalizationStates2}
1=\frac{1}{2} \langle \phi | \phi \rangle &=  \frac{1}{2} \int
\mathrm{d}^3 \vec{x} \, \langle \phi | x_+\rangle i
\on{\leftrightarrow}{\partial}_0 \langle x_+| \phi \rangle -
\langle \phi | x_-\rangle i \on{\leftrightarrow}{\partial}_0
\langle x_-| \phi \rangle\\ &= \frac{1}{2} \int \mathrm{d}^3 \,
\vec{x} \phi_+^{\ast}(x) i \on{\leftrightarrow}{\partial}_0
\phi_+(x) - \phi_-^{\ast}(x) i \on{\leftrightarrow}{\partial}_0
\phi_-(x) \nonumber \\ &= \frac{1}{2} \int
\frac{\mathrm{d}^3\vec{k}}{(2\pi)^3 2 \omega_k} \left[
|\phi(k_+)|^2 + |\phi(k_-)|^2 \right] \nonumber \\
&= \frac{1}{2} \int \frac{\mathrm{d}^4\vec{k}}{(2\pi)^3}
\delta(k^2+m^2) |\phi(k)|^2 \nonumber\,.
\end{flalign}
The momentum space realization of this inner product is discussed
for example in \cite{Morgan:2009wa} where the author considers a
stochastic interpretation of the fields. Halliwell and Ortiz
discuss in \cite{Halliwell:1992nj} both the position space and
momentum space realization in the context of the sum over
histories interpretation of quantum field theory. Their work was
inspired by earlier work of Henneaux and Teitelboim
\cite{Henneaux:1982ma}. In a slightly different form the inner
product can also be found in \cite{Woodard:1989ac}, also see
\cite{Mostafazadeh:2006fz}. The factor of $1/2$ that appears in
(\ref{NormalizationStates2}) is motivated by the canonical form of
the Hamiltonian in equation (\ref{Hamiltonian}). It is clear from
equation (\ref{NormalizationStates2}) that plane waves cannot be
normalized. One really needs a wave packet, i.e.: a superposition
of various wave modes with a certain amplitude, such that the
integrals in (\ref{NormalizationStates2}) converge. Note that a
single plane wave does not meet this requirement and is therefore
not a suitable wave packet. We can thus identify a probability
current:
\begin{equation}\label{Probabilitycurrent}
J^{P}_{\mu}(x) = \frac{1}{2} \left [\phi_+^{\ast}(x) i
\on{\leftrightarrow}{\partial}_\mu \phi_+(x) - \phi_-^{\ast}(x) i
\on{\leftrightarrow}{\partial}_\mu \phi_-(x) \right]\,,
\end{equation}
which is conserved for free theories in the usual sense:
\begin{equation}\label{Probabilitycurrent2}
\partial^{\mu} J^{P}_{\mu}(x) = 0\,,
\end{equation}
where we have used the Klein Gordon equation. We identify the 0th
component of the probability current as the probability density,
given a wave function $\phi(x)$:
\begin{equation}\label{Probabilitydensity}
\rho_P(x)= J^{P}_{0}(x) = \frac{1}{2} \left [\phi_+^{\ast}(x) i
\on{\leftrightarrow}{\partial}_t \phi_+(x) - \phi_-^{\ast}(x) i
\on{\leftrightarrow}{\partial}_t \phi_-(x) \right]\,.
\end{equation}
From the third line in equation (\ref{NormalizationStates2}) one
can clearly see that the probability density is both real and
positive. Given a spacelike region $\mathcal{A}$, the probability
of finding a particle in this region for some observer
thus follows as:
\begin{equation}\label{Probability}
P= \int_{\mathcal{A}} \mathrm{d}^3 \vec{x} J^{P}_{0}(x) =
\int_{\mathcal{A}} \mathrm{d}^3 \vec{x} \rho_P(x)=  \frac{1}{2}
\int_{\mathcal{A}} \mathrm{d}^3 \vec{x}  \left [\phi_+^{\ast}(x) i
\on{\leftrightarrow}{\partial}_t \phi_+(x) - \phi_-^{\ast}(x) i
\on{\leftrightarrow}{\partial}_t \phi_-(x) \right]\,.
\end{equation}
We already tentatively used the word particle in defining the
probability above. We make this statement more precise shortly.
Let us however first define the statistical propagator, or
Hadamard propagator as:
\begin{flalign}\label{HadamardPropagator1}
\Delta_H(x-x') &= i \Delta^+(x-x') - i \Delta^-(x-x') = - \langle
x_+| x_+' \rangle -  \langle x_-| x_-' \rangle \\
&=  - \int \frac{\mathrm{d}^4 k}{(2\pi)^3} \delta(k^2+m^2)
e^{ik(x-x')} = - \int \frac{\mathrm{d}^3 \vec{k}}{(2\pi)^3 2
\omega_k}\left[ e^{ik_+(x-x')} + e^{ik_-(x-x')}\right]\nonumber
\,.
\end{flalign}
We can straightforwardly relate the Hamadard propagator to the
imaginary part of the Feynman propagator:
\begin{equation}\label{HadamardPropagator3}
\Delta_H(x-x') = - 2  \Im\Delta_F(x-x') \,.
\end{equation}
The Hadamard propagator satisfies the following composition law in
terms of the commutator propagator:
\begin{equation}\label{compositionlaw7}
\int \mathrm{d}^3\vec{x}' \Delta_C(x-x'')
\on{\leftrightarrow}{\partial}_{\!t''}  \Delta_H(x''-x') =
\Delta_H(x-x')\,.
\end{equation}
The reason for introducing the Hadamard propagator now becomes
apparent. We can now show that:
\begin{equation}\label{HadamardPropagator2}
\frac{1}{2} \langle \phi | \phi \rangle =  \frac{1}{2} \int
\mathrm{d}^3\vec{x}\mathrm{d}^3\vec{x}' \left. \phi(x)
\on{\leftrightarrow}{\partial}_t \Delta_H(x-x')
\on{\leftrightarrow}{\partial}_{t'} \phi(x') \right |_{t=t'} =1
\,.
\end{equation}
This inner product can also be found in \cite{Halliwell:1992nj}.
The Hadamard propagator thus offers an alternative way of
evaluating the probability for finding a particle in our spacelike
region $\mathcal{A}$ for a particular observer:
\begin{equation}\label{Probability2}
P= \frac{1}{2} \int_{\mathcal{A}} \mathrm{d}^3 \vec{x}
\mathrm{d}^3 \vec{x}' \left. \phi(x)
\on{\leftrightarrow}{\partial}_t \Delta_H(x-x')
\on{\leftrightarrow}{\partial}_{t'} \phi(x') \right |_{t=t'} \,.
\end{equation}
The probability density $J^{P}_{0}(x)$ is not Lorentz invariant,
which is especially clear from the position space expression in
equation (\ref{Probabilitydensity}). The probability density
current $J^{P}_{\mu}(x)$ is, however, Lorentz covariant. Also, the
normalization of our probability to one, where $J^{P}_{\mu}(x)$ is
integrated over its entire domain, is Lorentz invariant. The
reason is that the inner product integral contains a Lorentz
covariant vector valued volume element, which contracts with the
current to give a Lorentz invariant result.

The probability current $J^{P}_{\mu}(x)$ in equation
(\ref{Probabilitycurrent}) merits a final remark. For real fields,
we have that $\phi_-(x) = \phi_+^{\ast}(x)$. Clearly, the
probability current is in that case equal to the ordinary Klein
Gordon current, given by:
\begin{equation}\label{KGcurrent}
N^{KG}_\mu(x) = \phi^*_+(x) i \on{\leftrightarrow}{\partial}_\mu
\phi_+(x)\,,
\end{equation}
where the negative frequency contributions in
(\ref{Probabilitycurrent}) are mapped to the positive frequency
Hilbert space. Therefore, equation (\ref{KGcurrent}) is a
probability density too. Such an interpretation can for example be
found in \cite{DeWit:1986it, Wald:1984rg, Brown:1992db}. Although
$J^{P}_{\mu}(x)$ and $N^{KG}_\mu(x) $ are equal, there are
profound differences between the dynamics of the wave function
$\phi(x)$, which is causal, and of $\phi_+(x)$, which is not
causal. Therefore, the positive frequency contribution to the real
field on its own cannot be a relativistic wave function.

\subsection{Particle Interpretation}

In non-relativistic quantum mechanics, a wave function describes
the probability density of a single particle. Motivated by the
discussion above, it is natural to interpret the probability
density in equation (\ref{Probabilitydensity}) as the number
density for a single particle too:
\begin{equation}\label{particlenumber}
N= \frac{1}{2} \langle \phi | \phi \rangle = \frac{1}{2} \int
\mathrm{d}^3 \, \vec{x} \phi_+^{\ast}(x) i
\on{\leftrightarrow}{\partial}_0 \phi_+(x) - \phi_-^{\ast}(x) i
\on{\leftrightarrow}{\partial}_0 \phi_-(x) =1\,.
\end{equation}
In other words, it is the superposition of positive and negative
frequency wave packets, normalized to unity, that defines one
particle. Hence, our particle interpretation is in agreement with
the particle interpretation as advocated in non-relativistic
quantum mechanics. Take for example the Gaussian ground state of a
simple harmonic oscillator. Clearly, it consists of the
superposition of many wave modes but is nevertheless generally
regarded as a single particle quantum mechanical wave function.
This differs from the particle interpretation that is usually
adhered to in quantum field theory, where one mode $|k\rangle =
\hat{a}_k^\dag |0\rangle$ is interpreted as one particle, see e.g.
\cite{Ryder:1985wq}. It also differs from various claims made in
the literature that the wave function for a single relativistic
particle does not exist in quantum field theory\footnote{We
\emph{are} able to adhere to a single particle wave function
interpretation since in our framework wave functions propagate
causally. As mentioned before however, there are still problems
related to non-locality in the standard treatments of position
operators. This issue will be resolved in \cite{Westra1}.}
\cite{Hegerfeldt:1974qu, Hegerfeldt:1998ar, Halvorson:2001hb,
Peres:2002wx, Haag:1992hx}.

The particle number in the sense defined above in equation
(\ref{particlenumber}) is only conserved in free theories. As for
the electric charge density, and for the probability density, we
can define a Lorentz covariant particle density current:
\begin{equation}\label{particledensitycurrent}
N_\mu = \frac{1}{2}\phi_+^*(x) i
\on{\leftrightarrow}{\partial}_\mu\phi_+(x) - \frac{1}{2}
\phi_-^*(x) i \on{\leftrightarrow}{\partial}_\mu \phi_-(x)\,.
\end{equation}
For complex fields for instance the charge current and the energy
momentum current are conserved because the Lagrangian following
from (\ref{Lagrangiandensity1}) is invariant under $U(1)$ and it
is translationary invariant. The particle number density current
defined in equation (\ref{particledensitycurrent}) is not
protected by such a symmetry in interacting theories.

\section{Causal Propagation of Quantum Fields}
\label{Causal Propagation of Quantum Fields}

\subsection{The Hadamard Composition Law and the Wheeler Propagator}

In the previous sections, we have developed a first quantized
picture for free scalar quantum field theories. In this final
section, we rigorously show that probabilities, defined in e.g.
equation (\ref{Probability2}) where we made use of the Hadamard
propagator, propagate causally.
\begin{figure}[t]
 \centering
  \begin{minipage}[t]{.95\textwidth}
   \includegraphics[width=\textwidth]{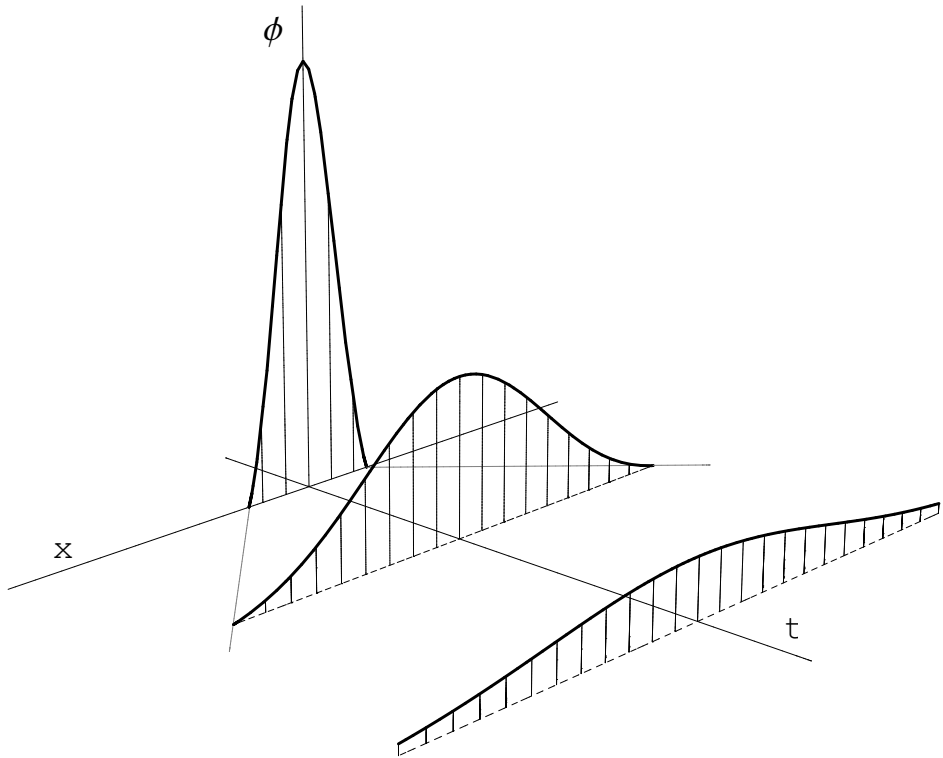}
   {\em \caption{Qualitative picture of the causal propagation of
   a quantum field mediated by the Wheeler propagator.
   Note that the evolution ensures that no contributions propagate
   beyond the causal region of the initial state. In other words, outside the causal future of the wave
   packet there is perfect destructive interference between the positive and negative
   frequency modes.
   \label{fig:causalpropagation} }}
   \end{minipage}
\end{figure}

Let us firstly remind the reader again that the Feynman propagator is
not a causal propagator. Suppose that, at some initial time $t'$,
one specifies a certain state $|\phi\rangle$, defined in equation
(\ref{fieldstates1a}). Let us now recall equation
(\ref{Feynmanpropagatora}):
\begin{equation}
\theta(t-t')\phi_+(x) = \int \mathrm{d}^3 \vec{x}'  \Delta_F(x-x')
\on{\leftrightarrow}{\partial}_{t'} \phi_+(x') \nonumber \,.
\end{equation}
The Feynman propagator has a non-vanishing support outside the
past and future light cone of $x'$ and hence we see that it
propagates the positive frequency contribution to our real field
$\phi$ instantaneously outside the future light cone of its region
of non-zero support. The Wheeler propagator stems from the
commutator, and thus only has non-vanishing support inside the
past and future light cone of $x'$. If we recall equation
(\ref{Wheelerpropagationlaw}):
\begin{equation}
\frac{\epsilon(t-t')}{2} \phi(x) =  \int \mathrm{d}^3 \vec{x}'
\Delta_W(x-x') \on{\leftrightarrow}{\partial}_{t'} \phi(x')
\nonumber \,,
\end{equation}
we conclude that $\phi(x)$ will only be non-zero if $x$ is the
past or future light cone of the region of non-zero support of
$\phi(x')$. Qualitatively, this is depicted in figure
\ref{fig:causalpropagation}.

The probability for some observer
for finding a particle in some
spacelike region $\mathcal{A}$, can be calculated by making use of
the Hadamard propagator, as we recall from equation
(\ref{Probability2}):
\begin{equation}
P= \frac{1}{2} \int_{\mathcal{A}} \mathrm{d}^3 \vec{x}
\mathrm{d}^3 \vec{x}' \left. \phi(x)
\on{\leftrightarrow}{\partial}_t \Delta_H(x-x')
\on{\leftrightarrow}{\partial}_{t'} \phi(x') \right |_{t=t'}
\nonumber \,.
\end{equation}
We can derive the following composition law for the Hadamard propagator:
\begin{equation}\label{compositionlaw8}
- \frac{1}{4} \epsilon(t-t_y) \epsilon (t'-t_{y'}) \Delta_H(x-x')
= \int \mathrm{d}^3\vec{y}' \mathrm{d}^3\vec{y}'' \Delta_W(x-y)
\on{\leftrightarrow}{\partial}_{\! t_y} \Delta_H(y-y')
\on{\leftrightarrow}{\partial}_{\! t_{y'}}\Delta_W(y'-x') \,.
\end{equation}
Here, we made use of equations (\ref{compositionlaw5}) and
(\ref{Wheelerpropagator2}). We have thus expressed the Hadamard
composition law in terms of the Wheeler propagator rather than the
commutator propagator as in equation (\ref{compositionlaw7}).
Also, we defined $y=(y_t,\vec{y})$. Alternatively, we can absorb
the minus sign in front of equation (\ref{compositionlaw8}) in the
composition law as follows:
\begin{equation}\label{compositionlaw9}
\frac{1}{4} \epsilon(t-t_y) \epsilon (t'-t_{y'})
\Delta_H(x-x') = \int \mathrm{d}^3\vec{y}' \mathrm{d}^3\vec{y}''
\Delta_W(x-y) \on{\leftrightarrow}{\partial}_{\! t_y}
\Delta_H(y-y') \on{\leftrightarrow}{\partial}_{\!
t_{y'}}\Delta_{AW}(y'-x') \,,
\end{equation}
where the anti-Wheeler propagator is defined as the anti-time
ordered commutator propagator:
\begin{equation}\label{antiWheeler}
\Delta_{AW}(x-x')=\frac{1}{2} \overline{T} \Delta_C(x-x') = -
\Delta_W(x-x') \,.
\end{equation}
If both $t$ and $t'$ are later than $t_y$ and $t_{y'}$, the
composition law (\ref{compositionlaw9}) takes a rather neat form:
\begin{equation}\label{compositionlaw12}
\Delta_H(x-x') = 4 \int \mathrm{d}^3\vec{y}' \mathrm{d}^3\vec{y}''
\left. \Delta_W(x-y) \on{\leftrightarrow}{\partial}_{\! t_y}
\Delta_H(y-y') \on{\leftrightarrow}{\partial}_{\!
t_{y'}}\Delta_{AW}(y'-x')\right|_{
\begin{subarray}{l}
\{ t,t'\} > t_y  \\
\{t,t'\} > t_{y'}
\end{subarray}}\,.
\end{equation}
We already showed before that the Wheeler propagator governs the
causal dynamics of the entire wave function $\phi(x)$. Now,
together with (\ref{Probability2}), equation
(\ref{compositionlaw12}) shows explicitly that probabilities
propagate causally too as the time evolution of the Hadamard
propagator can be expressed solely in terms of the causal Wheeler
propagator. A final technical remark is in order. The factor of
$4$ that appears on the right hand side in equation
(\ref{compositionlaw12}) arises from our definition of the Wheeler
propagator as a factor of half times the time ordered commutator
(\ref{Wheelerpropagator}), such that the Wheeler propagator is a
Green's function. If we would not have inserted the factor of
$1/2$ in our definition, the Hadamard propagator would satisfy a
neater looking composition law.

\subsection{Example I: a Localized Wave Function}

In order to illustrate the discussion above, let us study the
propagation of two wave packets: a specific $\phi_+(x)$ and
$\phi(x)$. We consider the following wave function $\phi$:
\begin{equation}\label{example1}
\phi(x) = \int \frac{\mathrm{d}^3\vec{k}}{(2\pi)^3}\frac{1}{2}
\left[ e^{i k_+(x-x')}+ e^{i k_- (x-x')} \right] \,.
\end{equation}
We have thus chosen the following wave packet:
\begin{equation}\label{example2}
\phi(k_\pm) = \omega_k e^{-i k_\pm x'} \,,
\end{equation}
The second wave packet we wish to consider is $\phi_+(x)$ which
can be read off from equation (\ref{example1}):
\begin{equation}\label{example3}
\phi_+(x) = \int \frac{\mathrm{d}^3\vec{k}}{(2\pi)^3}\frac{1}{2}
e^{i k_+(x-x')} \,.
\end{equation}
These wave packets do not give rise to a probability density that
can be normalized according to equation
(\ref{NormalizationStates}). The reason for introducing these wave
packets nevertheless is that they are perfectly localized
initially:
\begin{subequations}
\label{example4}
\begin{flalign}
\left. \phi(x) \right|_{t=t'} & =
\delta^{(3)}(\vec{x}-\vec{x}') \label{example4a} \\
\left. \phi_+(x) \right|_{t=t'} & = \frac{1}{2}
\delta^{(3)}(\vec{x}-\vec{x}')\label{example4b}\,.
\end{flalign}
\end{subequations}
These (extremely well) localized wave packets allow for a clean
comparison between the form of the wave packets for $t>t'$. The
conventional Newton-Wigner wave packet \cite{Newton:1949cq} is not
given by equation (\ref{example2}) but rather by:
\begin{equation}\label{example2b}
\phi_{+}^{\mathrm{NW}}(k) = \omega_k^{\frac{1}{2}} e^{-i k_{+} x'}
\,,
\end{equation}
and, consequently, gives rise to a wave function that is not
perfectly localized initially. In fact, the position space
expression for the Newton-Wigner wave packet yields at $t=t'$:
\begin{equation}\label{example3b}
\left. \phi_+^{\mathrm{NW}}(x)\right|_{t=t'} =
\frac{1}{2^{\frac{7}{4}} \pi^{\frac{3}{2}}
\Gamma\left(\frac{1}{4}\right)}
\left(\frac{m}{\|\vec{x}-\vec{x}'\|}\right)^{\frac{5}{4}}
K_{\frac{5}{4}}(m \|\vec{x}-\vec{x}'\|) \,,
\end{equation}
where $K_\nu(x)$ is a modified Bessel function of order $\nu$.
This expression reveals that the Newton-Wigner wave packet is not
perfectly localized initially.

All we need to do next, is perform the Fourier integrals in
(\ref{example1}) and (\ref{example3}). We make use of the form of
the two Wightman propagators (\ref{Wightmanfunctions}) and
introduce two $\epsilon$ pole prescriptions to regulate the
integral. Let us for simplicity consider a massless field such
that $\omega_k = \|\vec{k}\|$. Our wave function is given by:
\begin{flalign}
\phi(x) & = \zeta^{\mu}\partial_{\mu} \left[\Delta^+(x-x') +
\Delta^-(x-x')\right] = \partial_t \left[\Delta^+(x-x') +
\Delta^-(x-x')\right]  \label{example5}\\
&= \frac{\partial_t}{4\pi^2}\left[
\frac{i}{-(t-t'-i\epsilon)^2+\|\vec{x}-\vec{x}'\|^2} -
\frac{i}{-(t-t'+i\epsilon)^2+\|\vec{x}-\vec{x}'\|^2}\right]
\nonumber \\
& = \frac{i(t-t')}{2\pi^2}\left[
\frac{1}{\left[-(t-t'-i\epsilon)^2+\|\vec{x}-\vec{x}'\|^2
\right]^2} -
\frac{1}{\left[-(t-t'+i\epsilon)^2+\|\vec{x}-\vec{x}'\|^2 \right
]^2}\right] \nonumber\,.
\end{flalign}
In the first equality we have chosen our state such that
$\zeta^{\mu}$ is normal to the spatial hypersurface. Moreover, the
equations on the second and third line are to be interpreted in
the distributional sense where $\epsilon \rightarrow 0$. Most
importantly, our wave packet spreads in time and gives a
non-vanishing contribution only on the light cone. This is to be
contrasted with the result for $\phi_+(x)$ which results from
propagation forward in time with the Feynman propagator:
\begin{equation}\label{example6}
\phi_+(x)  = \partial_t \Delta^+(x-x')  =
\frac{i(t-t')}{2\pi^2\left[-(t-t'-i\epsilon)^2+\|\vec{x}-\vec{x}'\|^2\right
]^2} \,.
\end{equation}
This wave packet has non-vanishing contributions outside the light
cone too. We conclude that although initially both wave packets
are extremely well localized as delta functions, at later times
spreading of the wave packets has occurred such that our wave
function $\phi(x)$ has a non-vanishing contribution only on the
light cone whereas $\phi_+(x)$ has not propagated in such a causal
manner and has a non-vanishing contribution outside the light cone
too. This example thus shows that only $\phi(x)$ can be a wave
function consistent with the relativistic principle of causality.

\subsection{Example II: a Localized Probability Density}

We can also consider relativistic wave packets that can be
properly normalized according to equation
(\ref{NormalizationStates}). Let us now consider another wave
function $\phi(x)$ which is given by:
\begin{equation}\label{example7}
\phi(x) = \frac{1}{\sqrt{\mathcal{N}}} \int
\frac{\mathrm{d}^3\vec{k}}{(2\pi)^3 2 \omega_k} \left[ e^{i
k_+(x-x'-i \xi)}+ e^{i k_- (x-x'+i\xi)} \right] \,.
\end{equation}
Note that with this normalization $\phi(x)$ and $\mathcal{N}$ are
Lorentz scalars. This time, we have thus chosen the following wave
packet \cite{Kaiser:1990ww}:
\begin{equation}\label{example8}
\phi(k_\pm) =   \frac{1}{\sqrt{\mathcal{N}}} e^{-i k_\pm(x' \pm i
\xi)} \,,
\end{equation}
where $\xi$ is a timelike vector with $\xi^0>\|\vec{\xi}\|$ such
that the momentum space integral converges and where $ \mathcal{N}
$ is a normalization constant. For simplicity we will consider
$\xi^{\mu} = (\epsilon, 0,0,0) $ only. In particular, we will be
interested in the limit where $\epsilon$ is small. In principle,
however, we can generalize this analysis to more general states
where $\xi^0$ is arbitrarily large and where $\|\vec{\xi}\| \neq
0$ as long as the convergence condition $\xi^0>\|\vec{\xi}\|$ is
satisfied. Physically, states with $\|\vec{\xi}\| \neq 0$ have a
non-zero initial velocity of the wave packet. Also, $\xi^0$
determines the width of the wave packet. It is interesting to note
that for $m\neq0$, $x'=0$ and for small $\|\vec{k}\|$ the wave
packet in equation (\ref{example8}) is a Gaussian. For a more
general discussion of the dynamics of wave packets, see
\cite{Westra1}. Let us again specialize to massless scalar fields
for which $\omega_k = \|\vec{k}\|$. We can normalize our state by
making use of equation (\ref{NormalizationStates}):
\begin{equation}\label{example9}
\mathcal{N} = \frac{1}{(4\pi \epsilon)^2}\,.
\end{equation}
We can evaluate the Fourier integrals that appear in equation
(\ref{example7}) to find our wave function in position space:
\begin{equation}\label{example10}
\phi(x) = \frac{\epsilon}{\pi \left[-(t-t'-i \epsilon)^2+
\|\vec{x}-\vec{x}'\|^2 \right]} + \frac{\epsilon}{\pi
\left[-(t-t'+i \epsilon)^2+ \|\vec{x}-\vec{x}'\|^2 \right]} \,.
\end{equation}
We are interested in the probability density which we defined
earlier in equation (\ref{Probabilitydensity}):
\begin{equation}\label{example11}
\rho_P(x) = \frac{4\epsilon^3 \left[(t-t')^2+
\|\vec{x}-\vec{x}'\|^2 + \epsilon^2 \right]}{\pi^2 \left[-(t-t'-i
\epsilon)^2+ \|\vec{x}-\vec{x}'\|^2 \right]^2 \left[-(t-t'+i
\epsilon)^2+ \|\vec{x}-\vec{x}'\|^2 \right]^2} \,.
\end{equation}
As a simple computational check of the normalization of our
probability density, one can easily verify that:
\begin{equation}\label{example12}
\int \mathrm{d}^3\vec{x} \, \rho_P(x) = 1 \,,
\end{equation}
for arbitrary times as it should.
\begin{figure}[t!]
    \begin{minipage}[t]{.43\textwidth}
        \begin{center}
\includegraphics[width=\textwidth]{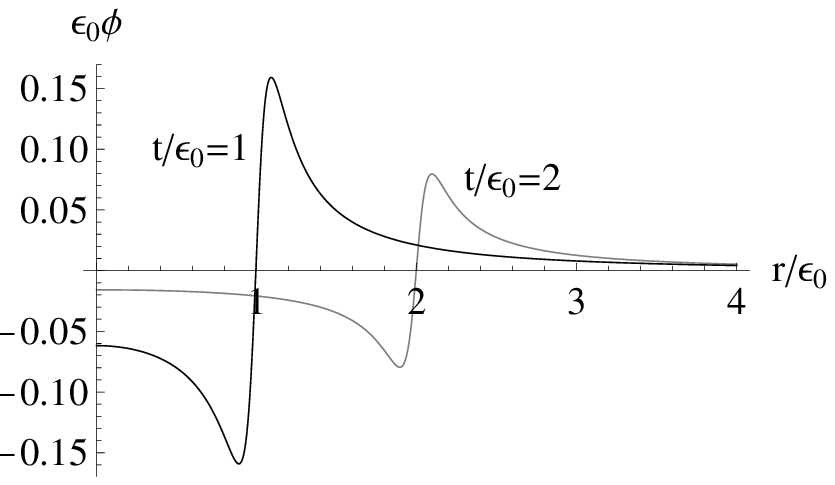}
   {\em \caption{The wave function given in equation
   (\ref{example10}) at $t/\epsilon_{0}=1$ and $t/\epsilon_{0}=2$. In
   non-relativistic quantum mechanics, it is hard to visualize the
   wave function as it is complex valued. Here, however, our wave
   function is the real Klein Gordon field. We used
   $\epsilon/\epsilon_0 =0.1$.
    \label{fig:wavefunction}}}
        \end{center}
    \end{minipage}
\hfill
\begin{minipage}[t]{.43\textwidth}
        \begin{center}
\includegraphics[width=\textwidth]{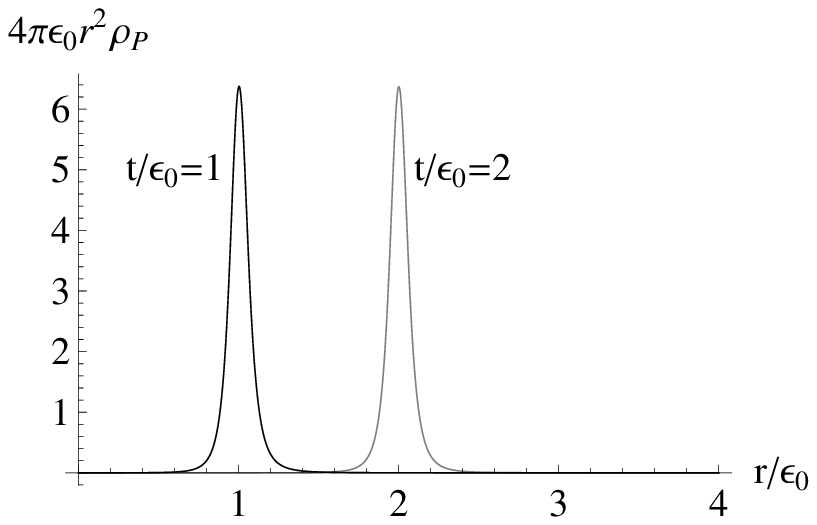}
   {\em \caption{The probability density in equation
   (\ref{example11}) resulting from the wave function depicted in figure
   \ref{fig:wavefunction}. Note that we multiplied by the surface of the
   spherical shell. We used $\epsilon/\epsilon_0 =0.1$.
    \label{fig:probabilitydensity}}}
        \end{center}
    \end{minipage}
\end{figure}
In figure \ref{fig:wavefunction} we show the real wave function in
equation (\ref{example10}) at two different times. In ordinary
quantum mechanics, one cannot so easily visualize the wave
function as it is complex valued. In figure
\ref{fig:probabilitydensity}, we show the resulting probability
density in equation (\ref{example11}). Note that it is a spherical
wave travelling at the speed of light. The reason for considering
a wave function of the form (\ref{example7}) is that initially its
corresponding probability distribution is extremely well
localized:
\begin{equation}\label{example13}
\left. \rho_P(x) \right|_{t=t'} =
\frac{4\epsilon^3}{\pi^2\left[\epsilon^2+  \|\vec{x}-\vec{x}'\|^2
\right]^3} \,,
\end{equation}
Sending $\epsilon$ to zero yields:
\begin{equation}\label{example14}
\lim_{\epsilon \rightarrow 0} \left. \rho_P(x) \right|_{t=t'} =
\frac{1}{2 \pi \|\vec{x}-\vec{x}'\|^2}\delta(
\|\vec{x}-\vec{x}'\|) = \delta^{(3)}(\vec{x}-\vec{x}') \,.
\end{equation}
Interestingly, the probability density for $t-t'>0$ is given by:
\begin{equation}\label{example15}
\lim_{\epsilon \rightarrow 0} \rho_P(x)  = \frac{1}{4 \pi
\|\vec{x}-\vec{x}'\|^2}\left[\delta \left(\|\vec{x}-\vec{x}'\| -
(t-t') \right) + \delta \left(\|\vec{x}-\vec{x}'\| + (t-t') \right
)\right] \,.
\end{equation}
For massless scalar fields, we thus observe that the probability
density propagates exactly \emph{on} the light cone. Furthermore,
it is important to realize that although $\phi_+(x)$ evolves with
the Feynman propagator and consequently spreads outside the light
cone of its initial region of non-zero support, the probability
density at late times behaves perfectly causal. The reason is that
our probability density given in equation
(\ref{Probabilitydensity}) and the Klein Gordon current in
equation (\ref{KGcurrent}) are equal. This is of course not
surprising. The reason for introducing the Wheeler propagator and
real valued relativistic wave functions has not been to prove
causality at the level of probabilities as expectation values in
quantum field theory are already perfectly causal. The main point
of our formalism, however, is that in-out amplitudes themselves
can be treated causally too. Finally, we would like to stress that
our single particle real valued relativistic wave function in
equation (\ref{example7}) is localized and can be properly
normalized. This is to be contrasted with the states constructed
by Newton and Wigner \cite{Newton:1949cq} as the eigenstates of
the Newton-Wigner position operator.

\section{Conclusion}
\label{Conclusion}

We have carefully developed a first quantized interpretation for
real scalar fields in non-interacting scalar field theory:
\begin{itemize}
\item The wave function of quantum field theory is the entire real
field $\phi(x)$, which generalizes the wave function of
non-relativistic quantum mechanics;
\item The dynamics of the wave function is governed by the Klein
Gordon equation, which generalizes the Schr\"odinger equation in
quantum mechanics;
\item The probability density follows from the innerproduct on the
Hilbert space of positive and negative frequency contributions to
$\phi(x)$. The probability density is of course normalized to
unity when integrated over its domain of non-zero support.
\end{itemize}
By making use of the stress-energy tensor for a real field we also
showed that the \emph{negative frequency} solutions have
\emph{positive energy}. The particle interpretation made in a
first quantized picture of quantum field theory differs from the
particle interpretation that is usually advocated in quantum field
theory. We interpret a wave function normalized to unity as the
quantum mechanical wave function of a single particle. Due to the
superposition principle, this wave packet thus contains many wave
modes. In conventional quantum field theory, one wave mode is
usually interpreted as one particle, see for instance
\cite{Ryder:1985wq}. The dynamics of the single particle wave
function $\phi(x)$ is governed by the Wheeler propagator, which is
the time ordered commutator propagator. The Wheeler propagator has
several advantages when compared to the Feynman propagator:
\begin{itemize}
\item Unlike the Feynman progagator, the Wheeler propagator is
causal. It does not have any contributions outside of the past and
future light cones of its arguments $x$ and $x'$;
\item The Wheeler propagator is real and propagates the entire real
field $\phi(x)$, while the Feynman propagator is complex valued
and only propagates the positive frequency contribution
$\phi_+(x)$;
\item Both the Wheeler and the Feyman propagator are
Green's functions, which makes them particularly convenient to
develop perturbation theory. However, only the Wheeler propagator
is the Green's function of the entire real field $\phi$.
\end{itemize}
Precisely because the Wheeler propagator or the commutator
propagator governs the causal propagation of the real field
$\phi(x)$, we can interpret the field itself as the wave function
in relativistic quantum mechanics. This is to be contrasted with
the function $\phi_+(x)$ that is virtually instantaneously
propagated outside the past or future light cone of its region of
non-zero support by the Feynman propagator.

In our approach the probabilities are calculated using the
Hadamard propagator. We only need the equal time Hadamard
propagator, so it plays no role in the dynamics of our scalar
field. The Hadamard propagator satisfies a composition law that
can be expressed solely in terms of the Wheeler propagator. We
have thus shown that probabilities in our form of the in-out
formalism propagate causally. In other words: causality is
restored in the in-out formalism of quantum field theory by
properly including the negative frequency solutions to the Klein
Gordon equation.

We would also like to mention again that expectation values of
observables in the first quantized picture of quantum field theory
will not be affected by
whether one makes use of the Feynman propagator or of the Wheeler
propagator to calculate them. So, standard quantum field theory in
the in-out formalism
is causal when expectation values of observables are considered,
but not when amplitudes are calculated. However, by making use of
the Wheeler propagator, we have been able to implement causality
explicitly at the level of the dynamics of the fields, unlike in
the conventional picture of first quantized quantum field theory.

The first quantized picture of quantum field theory we discussed
here is only complete if we can derive a local position operator.
This issue will be addressed in a separate publication
\cite{Westra1}. As said at various places throughout the paper, we
have only considered free real scalar fields. Of course, we are in
the process of generalizing our setup to interacting scalar fields
and set up a perturbative expansion using the Wheeler propagator
to calculate expectation values. Also, it would be very
interesting to generalize our setup to e.g. de Sitter spacetime.

\

\noindent \textbf{Acknowledgements}
\noindent The authors gratefully thank Tomislav Prokopec for many
useful comments and discussions. We also thank Thordur Jonsson and
Stefan Zohren for discussions at an earlier stage.


\end{document}